%% file: Rb_EPJD_v2.tex
\documentclass[epj]{svjour}
\usepackage[T1]{fontenc}
\usepackage[latin2]{inputenc}
\usepackage[english]{babel}
\usepackage[pdftex]{graphicx}
\usepackage{amssymb,amsmath}
\input{mathcommands.tex}
\DeclareMathOperator{\Tr}{Tr}

\begin{document}
\title{Photoionisation of Rubidium in strong laser fields}
\author{M.A.~Pocsai\inst{1,}\inst{2} \and I.F.~Barna\inst{1,}\inst{3} \and K.~T\H{o}k\'{e}si\inst{3,4}
}                     
%
%
\institute{Wigner Research Centre for Physics of the Hungarian Academy of Sciences 
\\ Konkoly--Thege Mikl\'os \'ut 29--33, 1121 Budapest, Hungary
\\ \email{pocsai.mihaly@wigner.mta.hu} \and University of P\'ecs, Institute of Physics, Ifj\'us\'ag \'utja 6 H-7624 P\'ecs, Hungary \and ELI-HU Nonprofit Ltd.\\
Dugonics T\'er 13, H-6720 Szeged, Hungary \and Institute for Nuclear Research, Hungarian Academy of Sciences\\
Bem square 18/c, H-4026, Debrecen, Hungary}
\date{Received: date / Revised version: date}
%
\abstract{
The photoionisation of rubidium in strong infra-red laser fields based on \textit{ab initio} calculations was investigated. The bound and the continuum states are described with Slater orbitals and Coulomb wave packets, respectively. The bound state spectra were calculated with the variational method and we found it reproduced the experimental data within a few percent accuracy. Using the similar approach, ionisation of Rb was also successfully investigated. The effects of the shape and the parameters of the pulse to the photoionisation probabilities and the energy spectrum of the ionised electron are shown. These calculations may provide a valuable contribution at the design of laser and plasma based novel accelerators, the CERN AWAKE experiment.
} 
\maketitle
\section{Introduction}
\label{intro}
The study of the ultrafast dynamics of electrons in intense short light pulses is a hot topic nowadays \cite{ultrafast}. Multiphoton photoionisation of one-, two- or many-electron atoms in intense laser field has been extensively studied by various non-perturbative quantum mechanical methods by different groups in the last decades \cite{diego-coulomb-volkov,fritzsche-pad,nagele-ishikawa,nagele-tdse,ishikawa}. One of them is the time-dependent close coupling (TDCC) method originally developed by Bray. The details can be found in the review of \cite{bray}. Another successful approach, the time-dependent close coupling method on a two-dimensional finite lattice was introduced by M.S.~Pindzola and F.~Robicheaux \cite{pindzola}. Various R-Matrix matrix methods are also feasible to perform ab-initio photoionisation calculations \cite{zatsarinny,hassouneh}.

As basis set expansions there are two popular ways. The first is the application of b-Splines by Bachau \cite{bachau} and the other is the Sturmian basis set \cite{randazzo}.

The problem of rubidium photoionizaton is almost ninety years old. The first experiment was performed by Lawrence \cite{lawrence-rb} where the wavelength of the applied light was in the range of 220 to 313 nm. Multiphoton ionizaton of Rb atoms was also studied with the help of a tunable dye-laser over the wavelength of 460 to 650 nm by Collins \cite{collins-rb}. Tamura and his group \cite{tamura} were the  first who measured the two-step selective photoionizatinon of Rb with Ti:sapphire solid state laser. Experimental determinaton of the photoelectron angular distribution of rubidium atoms in linearly and elliptically polarized lights were investigated by Wang and Elliott \cite{wang-elliot-rb}. Courtade {\it{et al}} calculated two-photon ionization of cold Rb atoms with a near resonant intermediate state. \cite{courtade-rb}.

Here we study the ionizaton of Rb atom problem with two different methods: full-fledged \textit{ab inito} quantum mechanical method and classical trajectory Monte Carlo (CTMC) method. In the former one, computationally we realized a time-dependent coupled-channel method  where the wave functions of the channels are constructed with Slater orbitals and regular Coulomb wave packets with equdistant finite widths in energy. This basis set was originally introduced to study heavy-ion and He atom collision in 2002 by Barna \cite{barnai} later become sophisticated and culminated to describe the angular distribution in two-photon double ionisation of helium by intense attosecond soft-x-ray pulses \cite{barnaiburgi}.

In this work we apply both TDCC and CTMC method to calculate the photoionisation probability of Rb interacting with strong laser field. We analyse how the photoionisation probabilities depend on the pulse parameters. We also calculate the photoelectron energy spectrum at different laser intensities with the TDCC method.

\section{Theory}
\label{sec:theory}

\subsection{\textit{Ab initio} calculations}
In our calculations we studied the photoionisation phenomena of rubidium via \emph{ab initio} calculations. The rubidium atom was modelled with a single active electron model with a frozen atomic core. Considering the time-dependent Schr\"{o}dinger-equation (TDSE):
\begin{equation}\label{eq:TDSE}
	i \partial_{t} \ket{\Psi (t, \mathbf{r})} = \hat{H} \ket{\Psi (t, \mathbf{r})}.
\end{equation}
The Hamiltonian operator has the form of
\begin{equation}
	\hat{H} = \hat{H}_{Rb} + \hat{V}_{I}
\end{equation}
where $\hat{H}_{Rb}$ and $\hat{V}_{I}$ describe the free rubidium atom and the interaction with the external laser field, respectively. There are many approaches to model the free rubidium atom, or in more general, an alkali atom within the confines of the one active electron approach \cite{green,garvey,garvey2}. When choosing such a model potential, the most important criterion is if the potential reproduces the experimental energy spectrum of the free atom. The Hellmmann pseudopotential, i.e.~the second term of \eqref{eq:Hamiltonian} fulfils this criterion and also has the advantage that many matrix elements can be calculated fully analytically. It also provides a graphical point of view to our physical approach: the inner shell electrons shield the electric charge of the atomic core, hence the valence electron feels a reduced charge from the core:
\begin{equation}\label{eq:Hamiltonian}
	\hat{H}_{Rb} = -\frac{1}{2} \nabla^{2} - \frac{1}{r}(1 - b e^{-dr}).
\end{equation}
The shielding parameters depend on the structure of a specific alkali atom. Ref.~\cite{milosevic} lists the parameters $b$ and $d$ for every alkali atoms. For rubidium, $b = 4.5$ and $d = 1.09993$.

The solution of the time dependent Schr\"{o}dinger equation can be expanded in terms of the eigenstates of the time-independent Schr\"{o}dinger equation:
\begin{equation}\label{eq:SE}
	\hat{H} \ket{\Phi_{j}(\mathbf{r})} = E_{j} \ket{\Phi_{j}(\mathbf{r})}.
\end{equation}
We apply the following Ansatz for the TDSE:
\begin{equation}\label{eq:ansatz}
	\ket{\Psi(t, \mathbf{r})}= \sum_{j=1}^{N} a_{j}(t)
\ket{\Phi_{j}(\mathbf{r})}e^{-iE_{j}t}.
\end{equation}
Inserting \eqref{eq:ansatz} into \eqref{eq:TDSE}, using \eqref{eq:SE}, we get:
\begin{equation}
	i \sum_{j=1}^{N}\dot{a}_{j}(t) \ket{\Phi_{j}(\mathbf{r})} e^{-i E_{j} t} = \sum_{j=1}^{N} \hat{V}_{I} a_{j}(t) \ket{\Phi_{j}(\mathbf{r})} e^{-i E_{j} t}.
\end{equation} 
Multiply the above equation by $\bra{\Phi_{k}(\mathbf{r})} e^{i E_{k} t}$. We get the following system of equations for the $a_{j} (t)$ coefficients:
\begin{equation}\label{eq:coupledchannels}
	\dot{a}_{k}(t) =  -i\sum_{j=1}^{N}V_{kj} e^{i E_{kj} t}a_{j}(t) 
\quad (k=1 \dots N).
\end{equation}
In \eqref{eq:coupledchannels}, $E_{kj} := E_{k} - E_{j}$ and $	V_{kj} := \bra{\Phi_{k}(\mathbf{r})} \hat{V_{I}} \ket{\Phi_{j}(\mathbf{r})}$
is the couplings matrix.

The (highly) oscillatory term can be transformed out by introducing
\begin{equation}\label{eq:a}
	\tilde{a}_{k}(t) := a_{k}(t) e^{-i E_{k} t}.
\end{equation}
Inserting \eqref{eq:a} into \eqref{eq:coupledchannels}, we get:
\begin{equation}\label{eq:coupledtransformed}
	i \dot{\tilde{a}}_{k}(t) = \sum_{j=1}^{N} V_{kj} \tilde{a}_{j}(t) + E_{k} \tilde{a}_{k}(t).
\end{equation}
In our calculations the initial conditions have been set up such that the valence electron is in the ground state:
\begin{equation}
	\begin{aligned}
a_{k} \left( t \rightarrow - \infty \right) =
\end{aligned}
\left\{
\begin{aligned}
& 1  \quad & k=1   \\
& 0 \quad & k \neq 1
\end{aligned} 
\right.
\end{equation}
We should note, however, that this is not a general requirement. Any mixed state could be specified as well. Integrating this system of ordinary differential equations on the time interval of the interaction, we get the wave function of the final state. Relevant physical quantities, e.g.~the occupation probabilities of either the bound or the continuum states can be calculated from this wave function:
\begin{equation}
	P_{k} (t \rightarrow \infty) = \abs{a_{k} (t \rightarrow \infty)}^{2}
\end{equation}
with $P_{k}$ denoting the occupation probability of the $k$-th state. The energy spectrum is defined by
\begin{equation}
	\frac{\partial P}{\partial E} = \sum_{l} \abs{\scalprod{\Phi_{E}^{l}(\mathbf{r})}{\Psi(t=T,\mathbf{r})}}^{2}
\end{equation}
with $\Phi_{E}^{l}(\mathbf{r})$ being an arbitrary continuum state with energy $E$ and azimuthal quantum number $l$ and $\Psi(t=T,\mathbf{r})$ being the final state wave function i.e.~the (numerical) solution of \eqref{eq:coupledtransformed}.

\subsubsection{Description of the bound and the continuum states}
As stated above, the wavefunction of the valence electron is expanded on the basis of the eigenfunctions of the free Hamiltonian operator.
It depends on the problem which basis is a sufficiently good choice. We describe the bound and the continuum states of the valence electron with Slater-type orbitals \eqref{eq:slater} and Coulomb wavepackets \eqref{eq:coulomb}, respectively \cite{barnai}:
\begin{equation}\label{eq:slater}
	\chi_{n,l,m,{\kappa}}(\vec{r}) 
= C(n,{\kappa})r^{n-1}e^{-{\kappa}r}Y_{l,m}    ({\theta},{\varphi})
\end{equation}
with $n,l,m$ being the principal, azimuthal and magnetic quantum numbers, respectively, and $\kappa$ the screening constant, which is, in our case, a variational parameter that specifies the energy of a bound state. The Slater orbitals form a normed, but not orthogonal basis, with the normalization factor
\begin{equation}
	C \parenth{n, {\kappa}} = \frac{\parenth{2 \kappa}^{n+1/2}}{\sqrt{\parenth{2n}!}}.
\end{equation}
The Coulomb wave packets discretize the continuum via integrating the Coulomb wave functions \eqref{eq:coulomb_wf} on a finite energy (momentum) interval. Therefore, they form the probability amplitude of an electron being in the state with its energy lying between $E - \Delta E$ and $E + \Delta E$. The energy and the width of such a state is given by the integration limits, i.e.~$k$ and $\Delta k$.
\begin{equation}\label{eq:coulomb}
	\varphi_{k,l,m,\tilde{Z}}(\vec{r}) = N(k,\Delta k) 
 \int\limits_{k-\Delta k/2}^{k+\Delta k/2} 
  F_{l,\tilde{Z}}(k',r)dk'Y_{l,m}({\theta},{\varphi})
\end{equation}
with $k$ and $\Delta k$ being the center and the width of the covered momentum range, $l$ and $m$ the azimuthal and magnetic quantum numbers, respectively, and $\tilde{Z}$ the charge of the ion, $\tilde{Z} = 1$ for a singly ionized atom. The Coulomb wave packets form an orthonormal basis such that if their corresponding energy ranges do not overlap, then their overlap integral is zero. The normalization factor reads:
\begin{equation}
	N(k, \Delta k) = \frac{1}{\sqrt{k \Delta k}}.
\end{equation}
Finally, the Coulomb wave function has the form of
\begin{equation}\label{eq:coulomb_wf}
\begin{split}
	F_{l,\tilde{Z}} (k,r) = &\sqrt{\frac{2k}{\pi}} \exp \parenth{\frac{\pi \tilde{Z}}{2k}} \frac{(2kr)^{l}}{(2l+1)!} \exp \parenth{-ikr} \times\\
	& \abs{\Gamma(l+1-i \tilde{Z}/k)} \times\\
	& _{1}F_{1}(1+l+i \tilde{Z}/k, 2l+2, 2ikr).
\end{split}
\end{equation}
For further details about the Coulomb wavefunctions, consult \cite{bethe-saltpeter}.

Sometimes, as it is in our case, it is more natural to specify the Coulomb wave packets with their energy range instead of the corresponding momentum range. There is a simple connection between the parameters of the momentum and the corresponding energy range:
\begin{subequations}
\begin{equation}
	k = \frac{1}{\sqrt{2}} \bparenth{\parenth{E + \frac{\Delta E}{2}}^{1/2} + \parenth{E - \frac{\Delta E}{2}}^{1/2}},
\end{equation}

\begin{equation}
	\Delta k = \sqrt{2} \bparenth{\parenth{E + \frac{\Delta E}{2}}^{1/2} - \parenth{E - \frac{\Delta E}{2}}^{1/2}}.
\end{equation}
\end{subequations}
We chose the parameters of the Coulomb wave packets such that they subdivide the total energy range equidistantly.

\subsubsection{Final formula for the energy spectra}
Determining the energy spectrum of the bound states leads to a variational problem. In general, the basis functions describing the bound states should contain at least one free parameter. In our case, every Slater orbital has a free parameter, the $\kappa$ screening constant, which will be determined via minimizing the energy functional corresponding to our physical system. By doing so, one gets a generalized eigenvalue problem:
\begin{equation}
	\mathbf{H} \mathbf{c} = E \mathbf{S} \mathbf{c}
\end{equation}
with
\begin{equation}
	H_{ij} = \sandwitch{\psi_{j}}{\hat{H}}{\psi_{i}}
\end{equation}
and
\begin{equation}
	S_{ij} = \scalprod{\psi_{j}}{\psi_{i}}
\end{equation}
being the Hamiltonian and the overlap matrices and $E$ the energy of a bound state, which is a generalized eigenvalue and $\mathbf{c}$ denotes a generalized eigenvector. Here $\psi_{j}$ can refer either to a Slater-function or to a Coulomb wave packet. Having $M$ states, an energy eigenstate of the free Hamiltonian operator with energy $E$ can be expanded on the chosen basis, the expansion coefficients being the components of the corresponding generalized eigenvector:
\begin{equation}
	\ket{\Phi_{j}(\mathbf{r})} = \sum_{p=1}^{M} c_{j,p} \ket{\psi_{p}(\mathbf{r})}.
\end{equation}

\subsubsection{Interaction with the external laser field}
In the configuration-interaction approach, introduced at the beginning of this section, every interaction with the external environment is incorporated into the couplings matrix. Its exact shape depends on the kind of the interaction, in our case, on the shape of the external laser field. Considering electromagnetic interaction, it is easy to show that the couplings matrix is proportional to the dipole matrix:
\begin{equation}
	V_{kj} = e \sandwitch{\Phi_{j}(\mathbf{r})}{\hat{V}_{I}}{\Phi_{k}(\mathbf{r})}.
\end{equation}
In length gauge, the interaction operator has the form of
\begin{equation}
	\hat{V}_{I} = \mathbf{r} \cdot \mathbf{E}(t, \mathbf{r}).
\end{equation}
Since in the present study we are investigating the interaction of a single atom and the external laser field, it is clear that the distances characterizing an atom are much smaller than the wavelength of the laser field. Therefore, the dipole approximation is valid. The electric field has the form of
\begin{equation}
	\mathbf{E}(t) = \boldsymbol{\varepsilon} E_{0} f(t)
\end{equation}
with $\boldsymbol{\varepsilon}$ being the polarization vector, $E_{0}$ the amplitude and $f(t)$ an arbitrary function that is feasible for modelling a laser pulse. The dipole matrix elements are therefore:
\begin{equation}
	D_{kj} = e \sandwitch{\Phi_{j}(\mathbf{r})}{\mathbf{r} \pmb{\varepsilon}}{\Phi_{k}(\mathbf{r})}\\
\end{equation}

Let $\mathbf{d}$ denote the dipole matrices corresponding the basis functions:
\begin{equation}
	d_{pq} := e \sandwitch{\psi_{q}(\mathbf{r})}{\mathbf{r} \pmb{\varepsilon}}{\psi_{p}(\mathbf{r})}.
\end{equation}
Using this notation, we get a compact formula for the dipole matrix elements:
\begin{equation}
	D_{kj} = \Tr \bparenth{\parenth{\mathbf{c}_{j}^{*} \circ \mathbf{c}_{k}} \mathbf{d}}.
\end{equation}
We chose the electric field such that it is polarized into the $z$ direction and has a sine-square envelope. The most common choice for the envelope is a Gaussian function, however, the sine-square envelope has a slight advantage at numerical calculations: such a pulse results in a function with compact support.
\begin{equation}
	\mathbf{E}(t) = \mathbf{e}_{z} E_{0} \sin^{2}\parenth{\frac{t}{T}} \sin \parenth{\omega_{L} t}
\end{equation}
with $T$ being the pulse duration and $\omega_{L}$ being the initial laser frequency.

\subsection{Classical trajectory Monte Carlo method}

A classical trajectory Monte-Carlo method is used to calculate the ionisation probabilities of Rb in intense laser fields. The Rb atom is characterised as two body system with an active electron and the remaining core. The interaction between the active electron and the target core is described either by a simple Coulomb potential as in the conventional CTMC method (model 1) or with the a model potential (model 2). In both versions of the present CTMC approach, Newton's classical non-relativistic equations of motions for a two-body system are solved numerically for a statistically large number of trajectories for given initial parameters. The equations of motion were integrated with respect to time as an independent variable by the standard Runge-Kutta method. Atomic units were used throughout the calculations. In this calculation, the total number of recorded trajectories was $1 \cdot 10^{5}$.

\subsubsection{Model 1}

The CTMC method is applicable to hydrogen-like target atoms (i.e., $A^{(z-1)+}$ ions, in general) in a natural manner. An extension to this picture requires an effective charge $Z_{\textrm{eff}}$ of the target nucleus as seen by the active electron:
\begin{equation}
\label{eq:screened-coulomb-potential}
	V \parenth{r} = -\frac{Z_{\textrm{eff}}}{r}.
\end{equation}
The effective charge of $2.2$ and corresponding binding energy of $0.154 \, \mathrm{a.u.}$ were used for Rb(5s).

\subsubsection{Model 2}
The potential of the Rb ion is represented by a central model potential developed by Green \cite{green} which was based on Hartree--Fock calculations:
\begin{equation}
\label{eq:green-potential}
	V \parenth{r} = - \frac{(Z-1) \Omega \parenth{r} + 1}{r}
\end{equation}
where $Z$ is the nuclear charge and
\begin{equation}
	\Omega \parenth{r} = \bparenth{H d \parenth{e^{r/d} - 1} + 1}^{-1}.
\end{equation}

Using the energy minimization, Garvey \textit{et al.}~\cite{garvey2} obtained the following parameters for Rb: H = $4.494 \, \mathrm{a.u.}$ and $d = 0.777 \, \mathrm{a.u.}$ The initialization parameters of Rb were selected as described by Reinhold and Falcon \cite{falcon} developed for non-Coulombic systems. The potential from Eq.~\eqref{eq:green-potential} is used to describe the interaction between the projectile and the target electron with the Rb ion core. The initial state of the target is characterized by a micro-canonical ensemble, which is constrained to an initial binding energy of $0.154 \, \mathrm{a.u.}$, at a relatively large distance from the collision center, choosing the initial parameters randomly. The distance between the projectile and the target was large enough that the interaction with the target was negligible.

\section{Results}
\label{sec:results}
In our calculations we modeled the rubidium atom by including the first 35 bound states. According to our the theoretical model we approximated the experimental energy levels by solving the corresponding variational problem that ended up in a generalized eigenvalue problem. By finding the proper values of the screening constants, the generalized eigenvalues of the Hamiltonian matrix corresponding to the bound states provide a sufficiently good approximation of the experimental energy levels. We managed to solve this optimization problem by using genetic algorithm (see e.g.~Ref.~\cite{genetic}). For the corresponding linear algebraic calculations we used the Armadillo library \cite{armadillo}. The calculated energy levels of the bound states, compared with the experimental are listed in tables \ref{tab:Ens}, \ref{tab:Enp}, \ref{tab:End} and \ref{tab:Enf}, respectively. The experimental data has been taken from \cite{nist}. $E_{\textrm{exp}}$ and $E_{\textrm{calc}}$ denote the experimental and the calculated energy eigenvalues, respectively.

Then, we defined an energy range in the continuum. We chose $0-15 \, \mathrm{eV}$ and subdivided it into 400 parts. The Coulomb packets with different azimuthal quantum numbers have been constructed to all energy levels accordingly, the azimuthal quantum numbers lying in the range $l = 0 - 3$ with $m = 0$. Since we included 35 bound states and there are 400 energy levels for every azimuthal quantum number in the continuum, we have a total of 1635 states in our model. For the first glance the energy resolution of the continuum spectra may seem to be too fine. However, during our calculations we saw that this is the resolution where the spectra converge to the same peak structures even at high intensities. That is, less fine resolution would have resulted in spectra with high numerical noise, whereas higher resolution would have been redundant. At this point we mention that choosing the proper energy resolution is a crucial point in similar calculations as the runtime of a single simulation grows rapidly as the intensity of the applied laser field grows. On the other hand, the runtime scales with $N^{2}$ with $N$ denoting the total number of states. At the highest applied intensity, using 20 processor cores took for almost two months.

\begin{table}
\begin{center}
\begin{tabular}{|r|l|l|}
\hline
State & $E_{exp} \, [a.u.]$ & $E_{calc} \, [a.u.]$ \\
\hline
  5s & -0.153507 & -0.148578 \\
\hline
 6s & -0.0617762 & -0.0649904 \\
\hline
 7s & -0.0336229 & -0.0370498 \\
\hline
 8s & -0.0211596 & -0.0236445 \\
\hline
 9s & -0.0145428 & -0.0161758 \\
\hline
 10s & -0.0106093 & -0.011617 \\
\hline
 11s & -0.00808107 & -0.00856655 \\
\hline
 12s & -0.00636018 & -0.00653549 \\
\hline
\end{tabular}
\caption{Energy spectrum of the bound states for $l = 0$. The experimental values have been taken from Ref.~\cite{nist}.}\label{tab:Ens}
\end{center}
\end{table}

\begin{table}
\begin{center}
\begin{tabular}{|r|l|l|}
\hline
State & $E_{exp} \, [a.u.]$ & $E_{calc} \, [a.u.]$ \\
\hline
 5p & -0.0961927 & -0.102753 \\
\hline
 6p & -0.0454528 & -0.0500378 \\
\hline
 7p & -0.0266809 & -0.0293907 \\
\hline
 8p & -0.0175686 & -0.0191912 \\
\hline
 9p & -0.0124475 & -0.0134602 \\
\hline
 10p & -0.00928107 & -0.00994177 \\
\hline
 11p & -0.00718653 & -0.00763469 \\
\hline
 12p & -0.00572873 & -0.00604301 \\
\hline
 13p & -0.0046738 & -0.00489973 \\
\hline
 14p & -0.0038856 & -0.00405129 \\
\hline
 15p & -0.00328125 & -0.00340454 \\
\hline
 16p & -0.00280771 & -0.00290064 \\
\hline
 17p & -0.00242976 & -0.00250062 \\
\hline
 18p & -0.00212316 & -0.00217684 \\
\hline
 19p & -0.0018712 & -0.00190818 \\
\hline
\end{tabular}
\caption{Energy spectrum of the bound states for $l = 1$. The experimental values have been taken from Ref.~\cite{nist}.}\label{tab:Enp}
\end{center}
\end{table}

\begin{table}
\begin{center}
\begin{tabular}{|r|l|l|}
\hline
State & $E_{exp} \, [a.u.]$ & $E_{calc} \, [a.u.]$ \\
\hline
 4d & -0.0653178 & -0.0544901 \\
\hline
 5d & -0.0364064 & -0.0307153 \\
\hline
 6d & -0.0227985 & -0.019709 \\
\hline
 7d & -0.0155403 & -0.0137109 \\
\hline
 8d & -0.0112513 & -0.0100841 \\
\hline
 9d & -0.00851559 & -0.0077216 \\
\hline
 10d & -0.00666683 & -0.00608392 \\
\hline
\end{tabular}
\caption{Energy spectrum of the bound states for $l = 2$. The experimental values have been taken from Ref.~\cite{nist}.}\label{tab:End}
\end{center}
\end{table}

\begin{table}
\begin{center}
\begin{tabular}{|r|l|l|}
\hline
State & $E_{exp} \, [a.u.]$ & $E_{calc} \, [a.u.]$ \\
\hline
 4f & -0.0314329 & -0.031225 \\
\hline
 5f & -0.0201073 & -0.0199783 \\
\hline
 6f & -0.0139554 & -0.0138734 \\
\hline
 7f & -0.0102476 & -0.0101937 \\
\hline
 8f & -0.00784234 & -0.00780359 \\
\hline
\end{tabular}
\caption{Energy spectrum of the bound states for $l = 3$. The experimental values have been taken from Ref.~\cite{nist}.}\label{tab:Enf}
\end{center}
\end{table}

We also took care of solving the coupled channel equations with the least numerical noise as possible. Since the pulse length is quite long, we decided to not to apply the common fourth order Runge--Kutta method when integrating our ODE system. Instead, we applied the Bulirsch--Stoer method with an eighth order embedded Runge--Kutta method as controller method. We set both the absolute and relative error tolerances to $10^{-8}$. This was sufficient---and also required---to preserve the unity norm of the wave function, i.e.~to not to numerically violate the unitarity. In our calculations the difference of the norm of the wave function from unity was not higher than $10^{-5}$.

After theses preparations, the spectrum and other quantities have been calculated such that the intensities lie in the range $10^{12} \, \mathrm{W}/\mathrm{cm}^{2} \leq I < 10^{14} \, \mathrm{W}/cm^{2}$ with $\lambda = 800 \, \mathrm{nm}$ wavelength and $\tau = 120 \, \mathrm{fs}$ pulse duration. The parameters have been chosen such that they fit to the CERN-AWAKE experiment \cite{awake}. Note that the Keldysh-parameter runs from $0.5916$ to $5.916$ if the intensities lie between $10^{14} \, \mathrm{W} \cdot \mathrm{cm}^{-2}$ and $10^{12} \, \mathrm{W} \cdot \mathrm{cm}^{-2}$.

\begin{figure}
\begin{center}
	\includegraphics[width=0.45\textwidth]{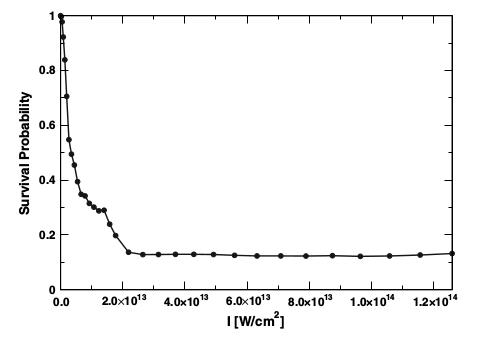}
	\caption{Stabilization of the potassium atom in a superatomic field: survival probability vs.~laser intensity for a $800 \, \mathrm{nm}$, $65 \, \mathrm{fs}$ laser pulse. For details, see ref.~\cite{morales}.}\label{fig:morales-stabilization}
\end{center}
\end{figure}

\begin{figure}
\begin{center}
	\includegraphics[width=0.45\textwidth]{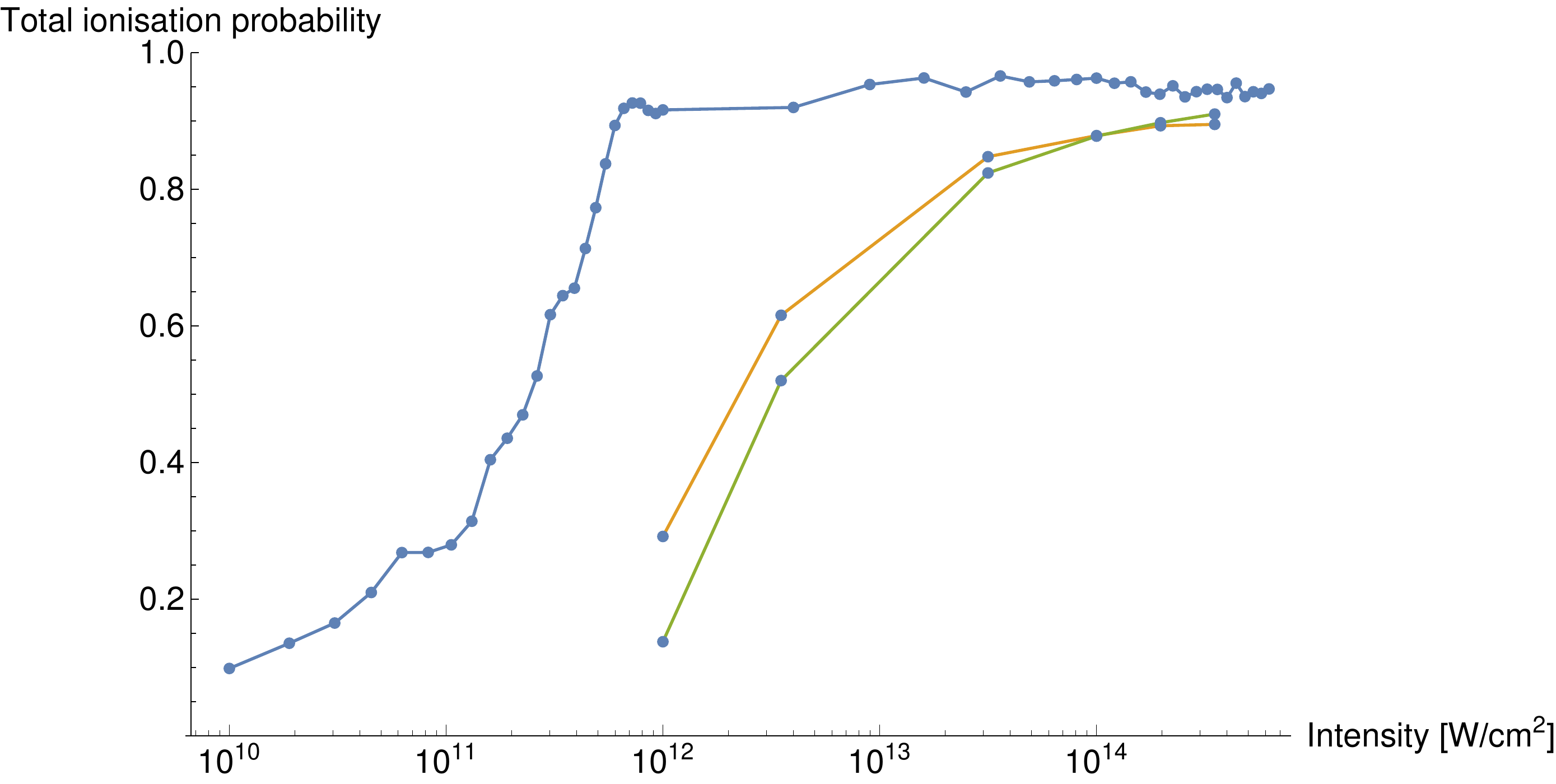}
	\caption{Total ionisation probabilities as a function of the laser intensity. $\lambda = 800 \, \mathrm{nm}$, $\tau = 120 \, \mathrm{fs}$. The blue line corresponds to the \textit{ab initio} calculations, the orange and the green to CTMC simulations with the model potentials \eqref{eq:screened-coulomb-potential} and \eqref{eq:green-potential}, respectively.}\label{fig:stabilization}
\end{center}
\end{figure}

According to Morales \textit{et.~al,} \cite{morales}, stabilisation of the total ionisation probability is expected (see Fig.~\ref{fig:morales-stabilization}).
This means that as the laser intensity increases, the total ionisation probability does not grow steadily, but instead, it either saturates at a given level, (Fig.~\ref{fig:morales-stabilization}), or reaches its maximum at a critical point, then drops slightly and remains constant at higher intensities (Fig.~\ref{fig:stabilization}). The critical point in the latter case is also referred as \textit{the bottom of the ``Death Valley''.}

Fig.~\ref{fig:stabilization} shows the photoionisation probabilities at different intensities with $\lambda = 800 \, \mathrm{nm}$ and $\tau = 120 \, \mathrm{fs}$. The blue line corresponds to the \textit{ab initio} calculations, the orange and the green to CTMC simulations with a classical Coulomb potential with Slater screening and Garvey model potential, respectively. According to the classical suppression \cite{reinold-ctmc,arbo-ctmc}, the CTMC simulations predict lower ionisation probabilities at lower intensities, but they also predict that the ionisation probability is going to be saturated as the laser intensity is growing. However, the discrepancy between the classical and quantum mechanical calculations is greater than expected. This difference could be corrected by fine-tuning the model potential used in the CTMC simulations. The \textit{ab initio} calculations suggest that the ionisation probability has been already saturated even at low intensities. This is a pleasant result since at the intensities planned to be applied in the CERN-AWAKE experiment practically fully ionisation occurs. This is a necessary condition for the experiment to work properly.

On Fig.~\ref{fig:stabilization} one can see that the classical suppression is larger than expected. Our idea is that this difference could be decreased by choosing a better model potential in the CTMC simulations

When should also note that in our case the stabilisation occurs at a much lower intensity than for potassium, even though the ionisation potentials of rubidium and potassium are quite similar ($4.177 \, \mathrm{eV}$ and $4.341 \, \mathrm{eV}$, respectively). The reason behind this that, in our model, the length of the laser pulse ($\tau = 120 \, \mathrm{fs}$) is almost the double of the one used in Ref.~\cite{morales} ($\tau = 65 \, \mathrm{fs}$). Therefore, the photon absorption is more ``efficient'' both in the bound states and the continuum states, resulting in a lower saturation intensity.

\begin{figure}
\begin{center}
	\includegraphics[width=0.45\textwidth]{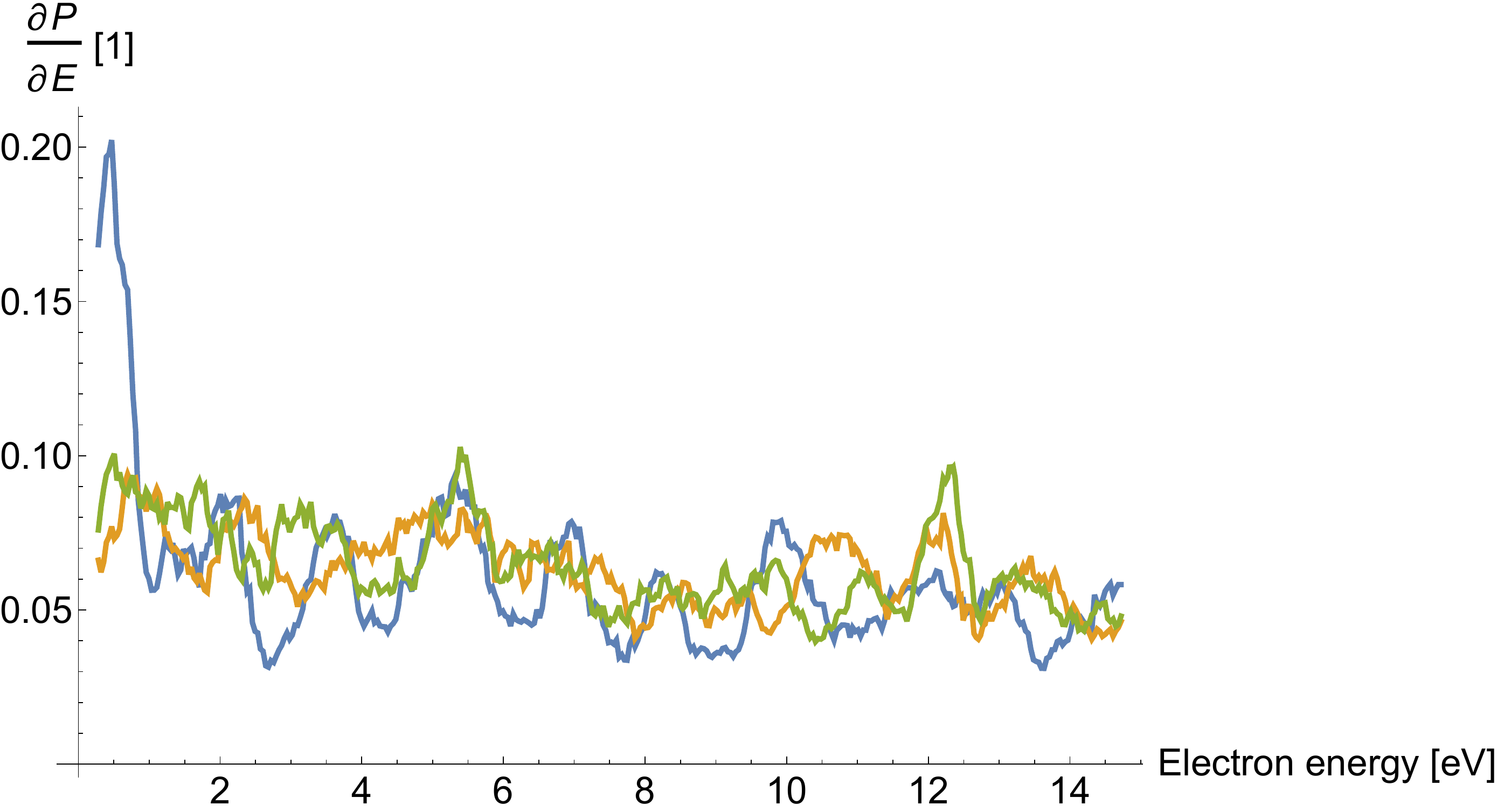}
	\caption{Energy spectrum of the continuum electron. The blue, orange and green lines are for $I = 10^{12}$, $2.5 \cdot 10^{13}$ and $8.1 \cdot 10^{13} \, \mathrm{W}/\mathrm{cm}^{2}$, respectively.}\label{fig:spectra}
\end{center}
\end{figure}

Photoelectron energy spectra have been calculated as well, see Fig.~\ref{fig:spectra}. Note that at relatively low intensities there is a strong peak near $0 \, \mathrm{eV}$. This means that at $10^{12} \, \mathrm{W}/\mathrm{cm}^{2}$ most of the electrons have relative low kinetic energy, that suggests that the plasma applied in the CERN-AWAKE experiment can have uniform electron density. At higher intensities none of the peaks is dominant, but all the spectra have a mutual characteristic. The difference between the abscissae of two neighbouring peaks is approximately $1.55 \, \mathrm{eV}$, i.e.~one photon energy at $\lambda = 800 \, \mathrm{eV}$ wavelength. This is a fingerprint of above-threshold ionisation (ATI) at these intensities. The width of these peaks originates from the non-resonant excitation during the ionisation process. Ref.~\cite{eberly-ati} provides an excellent review about this phenomenon.

\section{Conclusion}
In our work photoionisation of rubidium with strong laser pulses has been studied. We applied two different techniques: time-dependent close coupling and classical trajectory Monte Carlo method. In the former one, the continuum states have been described with Slater-type orbitals and the continuum states have been approximated with Coulomb wave packets. First we reconstructed the eigenfunctions of the free Hamiltonian operator on our basis set and found that the generalized energy eigenvalues of the Hamiltonian operator are in a good agreement with the experimental values. Then we investigated the dependence of the photoionisation probability and the photoelectron energy spectrum on the laser intensity. We chose the laser parameters such that they are similar to the corresponding ones in the CERN-AWAKE experiment. Since this parameter range is very specific, we could not directly compare our calculations with experimental data. However, our calculations reproduced the expected saturation of the photoionisation probability as a function of laser intensity.The \textit{ab initio} calculations predicted that staturation occurs at moderate intensities, near $10^{12} \, \mathrm{W} / \mathrm{cm}^{2}$, whilst CTMC simulation predicted this phenomena only near $10^{14} \, \mathrm{W} / \mathrm{cm}^{2}$. Both methods agree qualitatively with each other and also with the presented simulation of Morales \textit{et al.}~\cite{morales}. We also calculated the photoelectron energy spectrum at different laser intensities. The ATI peaks are clearly visible and it can also be seen that the average kinetic energy of the continuum electron grows monotonically with the laser intensity.

For a better understanding of the physics of photoionisation of Rb further work is in progress.
We plan to calculate photoelectron angular distributions and to investigate the additional effects regarding the variation of additional laser parameters like pulse duration or wavelength. We also plan to study the phenomena related to non-zero chirp.

\section{Acknowledgement}
The ELI-ALPS project (GINOP-2.3.6-15-2015-00001) is supported by the European Union and co-financed by the European Regional Development Fund.One of us K. T\H{o}k\'{e}si acknowledges the support by the National Research, Development and Innovation Office (NKFIH), Grant No. KH 123886. Author M.~A.~Pocsai would like to acknowledge the support by the Wigner GPU Laboratory of the Wigner RCP of the H.A.S.

\section{Statement}
This article is a basis of Author M.A.~Pocsai's Ph.~D Thesis. He is responsible for preforming and evaluating the \textit{ab-inito} calculations, including implementing the corresponding software, written in C++11. He is primarily responsible for writing the manuscript. Author I.F.~Barna is the supervisor of M.A.~Pocsai. He also participated in writing and reviewing the manuscript. Author K.~T\H{o}k\'{e}si is primarily responsible for performing and interpreting the CTMC calculations that have been compared with the \textit{ab-initio} results. Furthermore, he participated in writing and reviewing the manuscript.

\bibliographystyle{epj}
\bibliography{citations}

\end{document}

%% file: mathcommands.tex
\newcommand{\parenth}[1]{\left( #1 \right)}
\newcommand{\bparenth}[1]{\left[ #1 \right]}

\newcommand{\abs}[1]{\left\lvert #1 \right\rvert}

\newcommand{\ket}[1]{\left\lvert \left. #1 \right\rangle \right.}
\newcommand{\bra}[1]{\left\langle \left. #1 \right\rvert \right.}
\newcommand{\scalprod}[2]{\left\langle #1 \vert #2 \right\rangle}
\newcommand{\sandwitch}[3]{\left\langle #1 \abs{#2} #3 \right\rangle}

